\documentclass[a4paper,11pt]{article}
\usepackage{jinstpub} % for details on the use of the package, please see the JINST-author-manual
\usepackage{lineno}
\usepackage{siunitx}
\usepackage{subcaption}
\title{\boldmath Test-Jig for production testing of RPC-DAQ modules in INO-ICAL Experiment}

\author[a,1]{Yuvaraj Elangovan,\note{Corresponding author.}}
\author[a]{Mandar Saraf,}
\author[a]{B. Satyanarayana,} 
\author[a]{S.S. Upadhya,} 
\author[a]{Ravindra Shinde,} 
\author[a]{Gobinda Majumder,} 
\author[b]{Purnendu Kumar,}
\author[a]{S. Thoi Thoi,}
\author[a]{and Aditya Deodhar}
\affiliation[a]{Tata Institute of Fundamental Research, Mumbai, India}
\affiliation[b]{Indian Institute of Technology Madras, Chennai, India}

\emailAdd{yue8@pitt.edu}

\abstract{
The INO-ICAL experiment consist of 28,800 RPCs each equipped with a Front-End FPGA-based Data Acquisition (RPC-DAQ) module for acquiring detector signals. Due to the large number of RPC-DAQs are required, an automated test system is essential. RPC-DAQ Test-Jig is an FPGA module designed to generate standard test inputs to the RPC-DAQ supporting complete functionality testing. The RPC-DAQ has multiple functions such as strip hit latching, count rate monitoring, pulse stretching, trigger generation, TDC data collection, and Ethernet communication. To effectively test each of these logic's the Test-Jig uses various test patterns allowing users to verify and debug RPC-DAQ modules at a faster rate. When generating a predefined event with known data the Test-Jig architecture generates data pattern similar to that of the detector and also verifies the received data from the RPC-DAQ simultaneously. This testing methodology helps in understanding the functionalities of the RPC-DAQ logic at various conditions. The developed Test-Jig and along with its test methodologies reduces the debugging time of RPC-DAQs. This Paper discuss the architecture of the Test-Jig and some of its test methodologies in detail.}


\begin{document}
\maketitle
\flushbottom
% ~\ref{fig:2}, ~\ref{tab:2} , ~\cite{l}
\section{Introduction}
\label{sec:intro}
The India based Neutrino Observatory aimed at studying various properties of rarely interacting particle neutrinos. INO ICAL experiment~\cite{a} plans to construct a 51-kiloton magnetized iron calorimeter for the investigation of atmospheric neutrinos. In ICAL 28,800 Resistive Plate Chambers (RPCs) will be positioned inside the ICAL magnet.

RPCs are gaseous detectors created by sandwiching two glass plates with polycarbonate spacers maintaining a uniform $2\, \text{mm}$ gap throughout the plane. Ionization occurs in the gas volume when a charged particle traversing through them. Ionized electron moves due to the external electric field and multiples by a factor of \num{e+6}. The motion of those multipled electrons induce signals on pickup panels placed orthogonally on top and bottom of the gas chamber. These pickup panel strips connect to the Analog Front End (AFE) responsible for further amplifying and discriminating signals. Each RPC is equipped with its dedicated data acquisition system (RPC-DAQ)~\cite{b}, receiving 128 strip signals in LVDS standard from the AFE. These signals carry position and timing information of incident particle.

\begin{figure}[htbp]
\centering
\includegraphics[width=.5\textwidth]{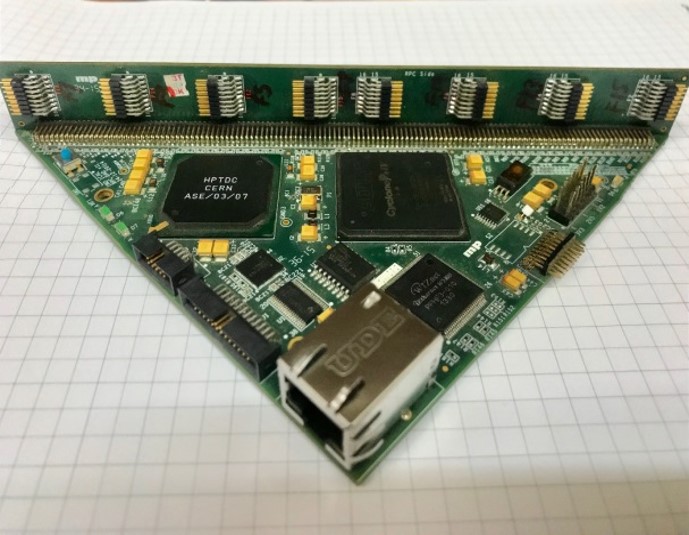}
\caption{RPC-DAQ Module. \label{fig:1}}
\end{figure}

The RPC-DAQ module, depicted in Figure ~\ref{fig:1} comprises of a Cyclone 4 Intel FPGA for signal processing. A soft-core processor NIOS~\cite{c} is instantiated in the FPGA for offloading data transfer. Also a $100\, \text{ps}$ resolution time-to-digital converter for timestamping is integrated. An Ethernet interface is used for data transfer between RPC-DAQ and Back-end using the standard TCP/IP protocol. Major functionalities of RPC-DAQ includes latching of strip signals on a global event trigger, accruing timing information, monitoring detector health, timestamping with a RTC clock of resolution $100\, \text{ns}$ and data transportation over Ethernet. These functionalities are developed to handle detector inputs at various boundary conditions. So it is important that the logic implemented in the RPC-DAQ functions properly to avoid issues during run time. Additionally the integrity of the RPC-DAQ module needs thorough testing before deployment. 

 \begin{figure}[htbp]
\centering
\includegraphics[width=.8\textwidth]{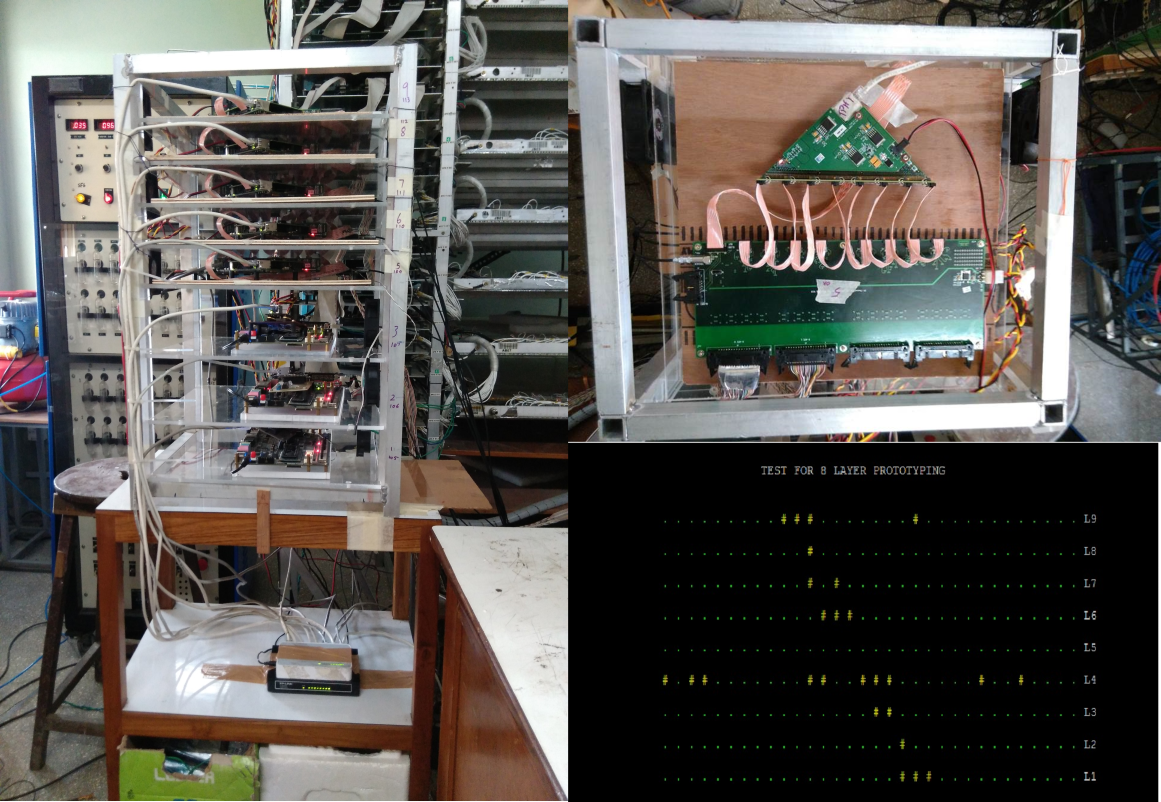}
\caption{RPC-DAQ testing using Prototype INO RPC Detector Stack and Result of track observed.  \label{fig:2}}
\end{figure}

An initial set of RPC-DAQs was produced and interfaced with a prototype 12-layer INO RPC stack ~\cite{d} located at TIFR, Colaba, Mumbai, India. During testing, RPC-DAQs were connected to $1\, \text{m} \times 1\, \text{m}$ RPCs, each consisting of 32 channels on the X-side and 32 channels on the Y-side. An 8-layer DAQ stack as shown in Figure~\ref{fig:2} was developed and placed close to the RPC stack, with part of the AFE signals connected to the RPC-DAQ.  A 64-channel ECL to LVDS adapter board was designed to act as a level translator between the detector front-end DFE and RPC-DAQ. A global trigger was copied from the VME-based DAQ system which controls the overall data acquisition of the 12 layer RPC stack.
 
 \begin{figure}[htbp]
\centering
\includegraphics[width=.8\textwidth]{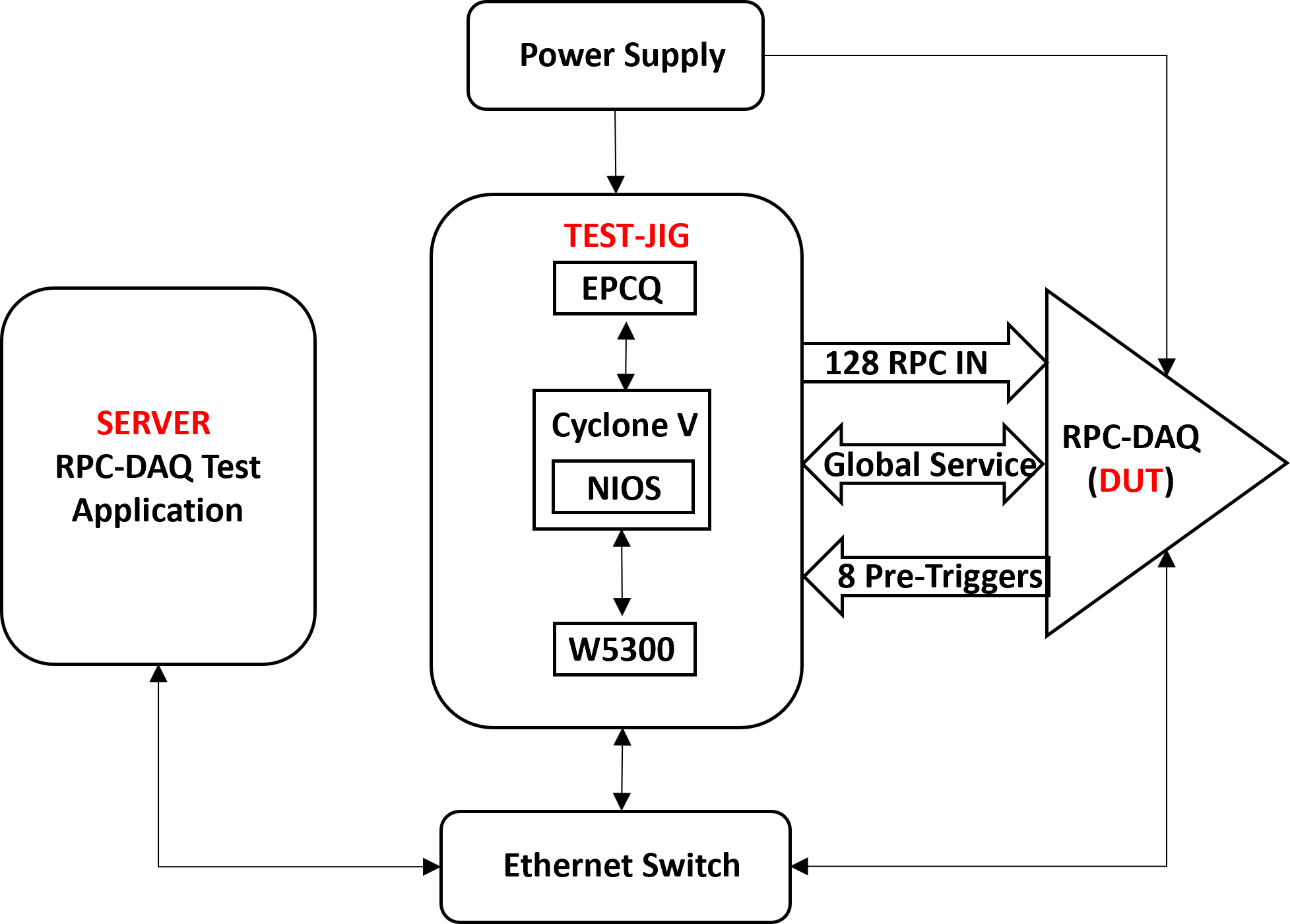}
\caption{RPC-DAQ Test Architecture.\label{fig:3}}
\end{figure}
It's important to note that the event triggers are generated in the VME based data acquisition system and tapped by RPC-DAQs to collect events. This approach allows RPC-DAQs to be tested in a realistic environment with event triggers. The resulting data from RPC-DAQs are collected over Ethernet in a remote server. A basic run control software based on C++ was developed for this purpose. The collected events are carefully analysed to check the functionality of RPC-DAQ hit latching and plotted as shown in Figure~\ref{fig:2}. An automated testing system which generates realistic inputs is required to test and debug 28,800 RPC-DAQ modules at a fast rate. To explore this process a test architecture has been designed as illustrated in Figure~\ref{fig:3}.

This involves generating and passing LVDS signals similar to those from AFEs using another FPGA module. This paper presents a basic test architecture which uses an FPGA based module and variety of test methods for the large-scale testing of RPC-DAQs.

\section{Test-Jig Module Description }
\label{sec:proto}

\begin{figure}[htbp]
\centering
\includegraphics[width=.8\textwidth]{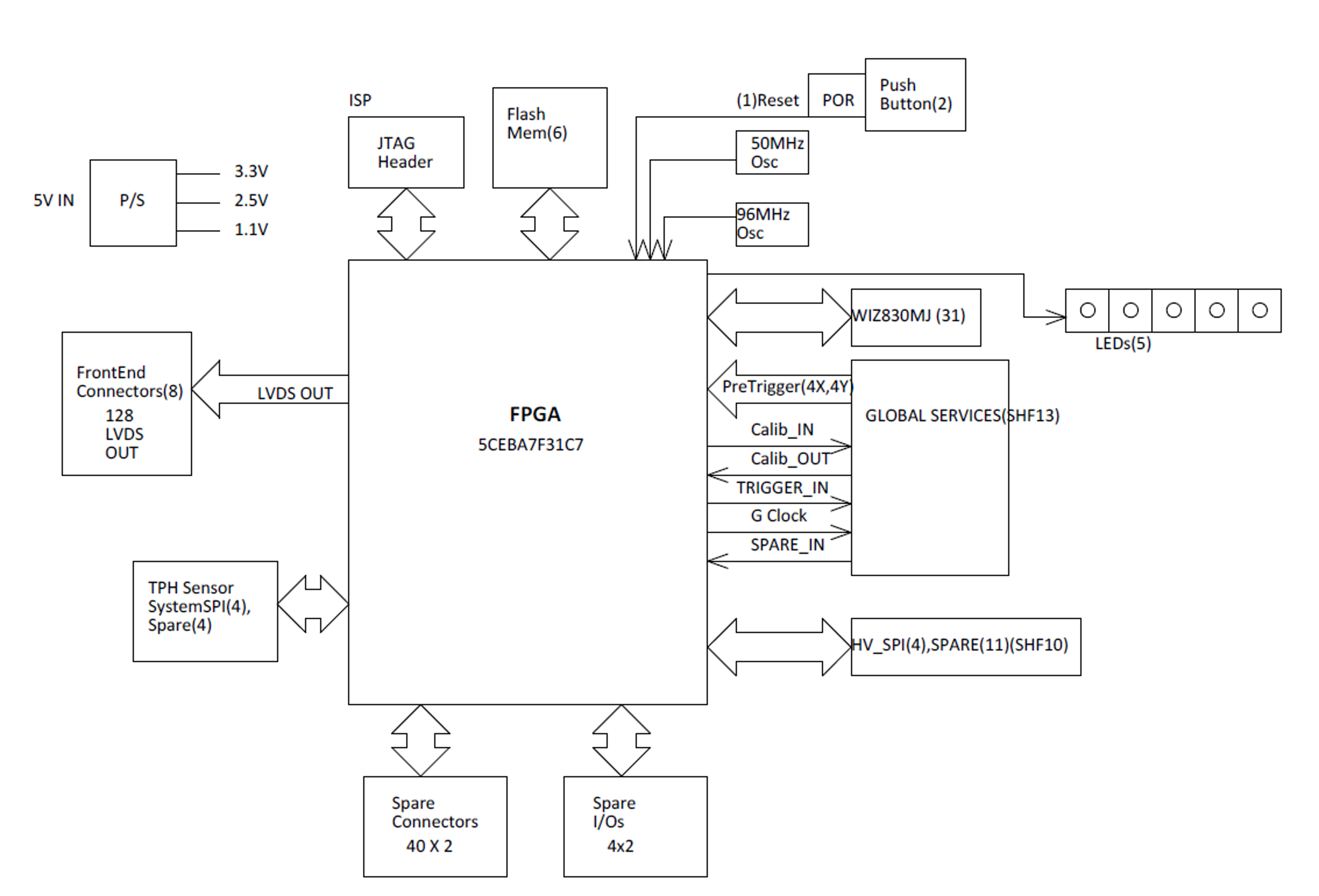}
\caption{Testjig Block diagram. \label{fig:4}}
\end{figure}

For this purpose, a Test-Jig module based on Cyclone 5 FPGA ~\cite{e} was developed. The Test-Jig, depicted in the block diagram Figure~\ref{fig:4}, emulates the interface experienced by RPC-DAQ in real-world scenarios.

The Test-Jig is equipped with an Ethernet interface to process test commands and generate test patterns. An integrated application software, RPC-DAQ Test Application (RTA), was developed to control the Test-Jig and function as an RPC-DAQ data server. Specific test methodologies were created for testing each RPC-DAQ functionality. The Test-Jig firmware was designed to handle all these methodologies, as dictated by RTA, and permanently programmed into the Test-Jig.

The Test-Jig module can interface with only one RPC-DAQ at a time. A common Ethernet switch connects the Test-Jig, RPC-DAQ and RTA server. During large scale testing multiple Test-Jigs will be connected to RPC-DAQs in parallel to accelerate the testing rate. The RTA can simultaneously test multiple RPC-DAQs with an artificial physics event provided there are more than 10 Test-Jig-RPC-DAQ setups available. This method can be used for testing the group functionality of RPC-DAQs such as acquiring a physics event with global timestamps and also assesses other subsystems of ICAL including trigger and calibration systems.
\begin{figure}[htbp]
\centering
\includegraphics[width=.8\textwidth]{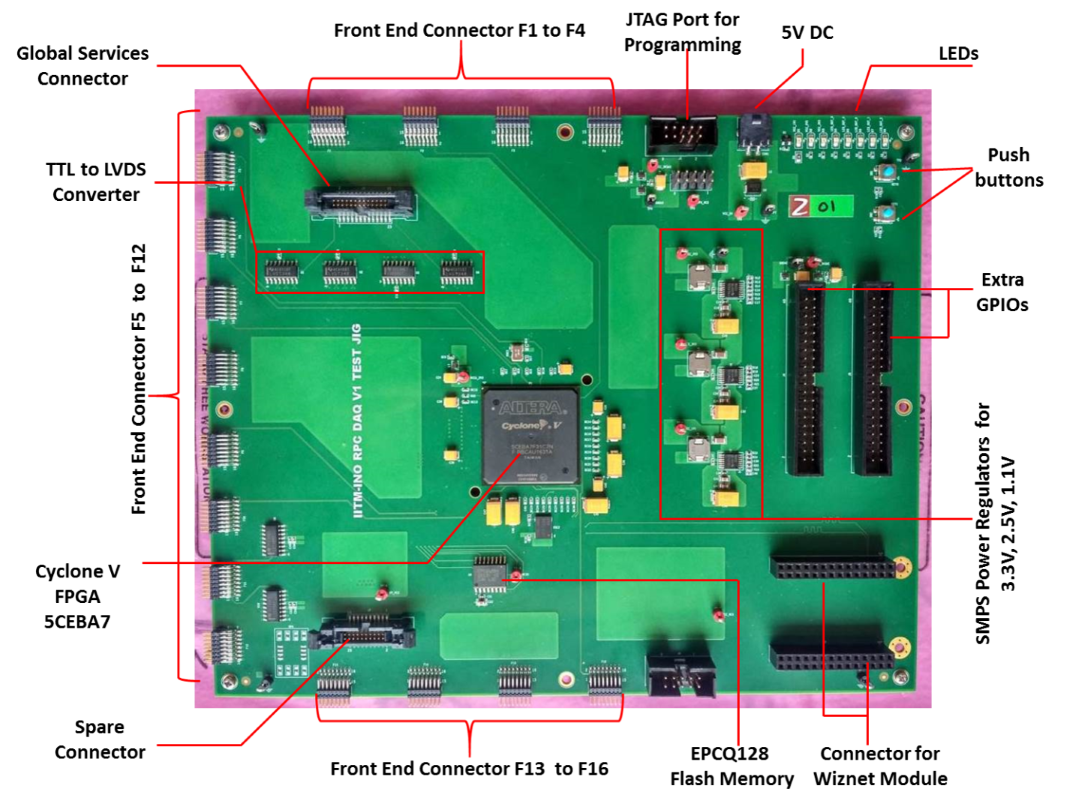}
\caption{Fabricated Test-Jig Module.\label{fig:5}}
\end{figure}
The Test-Jig module shown in Figure~\ref{fig:5} is an 8-layer PCB with dimensions of $220\, \text{mm} \times 180\, \text{mm}$ and an overall thickness of $1.6\, \text{mm}$. At the core of the Test-Jig is a Cyclone 5 FPGA with $150\, \text{k}$ logic elements and 480 GPIOs. A dedicated 128 Mbit Flash memory (EPCQ128)~\cite{f} is interfaced with the FPGA for configuration and it can also be utilized to store test patterns. Two crystal oscillator clock sources (50 MHz and 96 MHz) are connected to the FPGA as system clocks. The Wiznet W5300~\cite{g} an Ethernet controller helps in offloading FPGA data transportation featuring a complete Ethernet TCP/IP stack spread over 8 sockets. Each socket can be configured by the user for different protocols (UDP/TCP) and it includes 128 KB memories for transmit and receive buffers covering all 8 sockets.

All additional ports available in the FPGA are connected to spare connectors. For connecting global services to RPC-DAQ a dedicated connector is interfaced with the FPGA via 16 TTL to LVDS converters. These global services interface provides a 10 MHz clock, a Pulse Per Second (PPS) clock and trigger to RPC-DAQ. It is also utilized to read the trigger primitive that is fold signals from RPC-DAQ including 1fold, 2fold, 3fold, and 4fold for both X and Y signals. The Test-Jig has a maximum power rating of 3 Watts and operates on a 5V DC power supply. The Test-Jig FPGA can drive a maximum of 120 LVDS output channels and the remaining 8 channels are driven as TTL from the FPGA converted to LVDS using TTL to LVDS buffers outside the FPGA.

\section{Test-Jig RPC-DAQ Test Setup}
\label{sec:conf}

\begin{figure}[htbp]
\centering
\includegraphics[width=.8\textwidth]{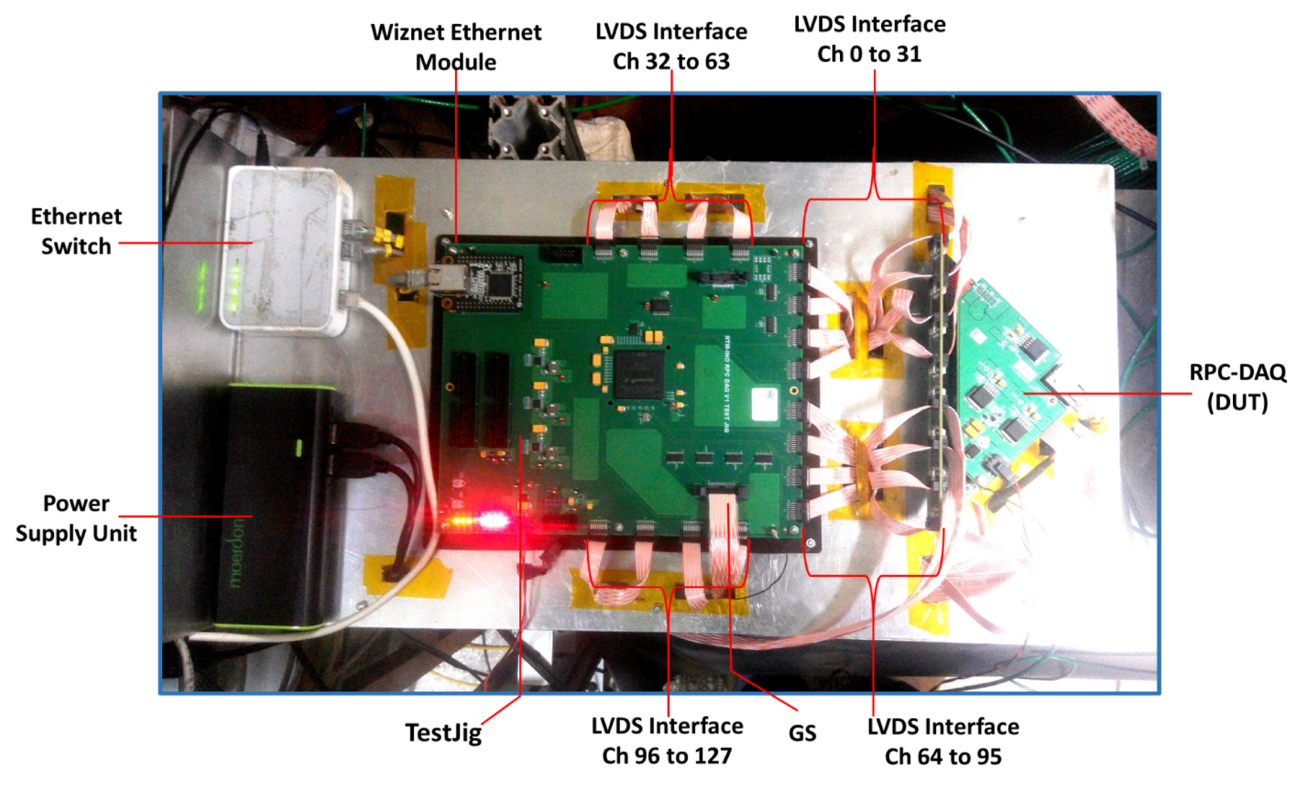}
\caption{Test-Jig Module Interface with RPC-DAQ. \label{fig:6}}
\end{figure}

The Test-Jig module was integrated with the RPC-DAQ as illustrated in Figure~\ref{fig:6}. To facilitate this setup a $500\, \text{mm} \times 280\, \text{mm}$ aluminum chassis was prepared. Small rectangular slits were cut for efficient cable routing between the Test-Jig and all interfaces. The RPC-DAQ was mounted adjacent to the Test-Jig to establish connections with the LVDS channels. 

A dedicated Ethernet switch was installed in the chassis to distribute Ethernet connectivity to the Test-Jig and RPC-DAQ from the server. Both RPC-DAQ and Test-Jig share common 5V DC power supplies. High-density twisted pair cables with a $1.27\, \text{mm}$ pitch were used for LVDS connections between the Test-Jig and RPC-DAQ. One notable challenge is that each time the RPC-DAQ needs to be positioned and connected with all the cables before initiating the test.

\section{RPC-DAQ Test Application (RTA)}
\label{sec:flow}
\begin{figure}[htbp]
\centering
\includegraphics[width=.8\textwidth]{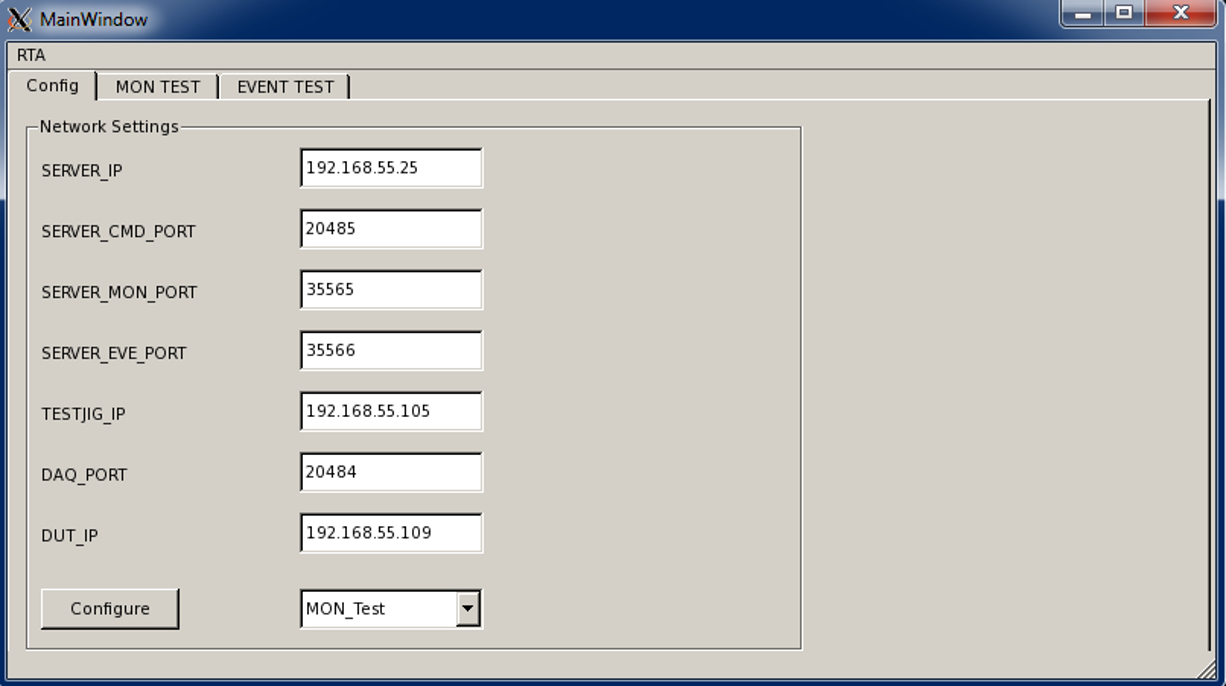}
\caption{RTA software Graphical User Interface.\label{fig:7}}
\end{figure}

RTA is PyQT4 based application software. It runs on both Linux and windows platforms. Basically it contains a connection less UDP socket program to send and receive commands with Test-Jig and RPC-DAQ. Also it has separate TCP connection based socket threads for receiving real time event data from RPC-DAQ. The command interface implemented uses necessary handshakes to ensure command integrity. The UDP command interface used by RTA works on two modes MULTICAST and UNICAST. Multicast mode is for common commands for both RPC-DAQ and Test-Jig whereas Unicast mode is used for sending commands only to specific modules either Test-Jig or RPC-DAQ. Using a separate configuration tab user can configure the device under test RPC-DAQs hardware address like IP address and Port numbers as shown in Figure~\ref{fig:7}. 

Upon receiving user instructions the RTA selects a test methodology and generates an input data pattern. This data is formatted into a UDP packet and transmitted to the Test-Jig as a Unicast packet. Also the RTA retains a copy of the generated data pattern. These data are transferred to dedicated RTA threads for comparison with the incoming data from the RPC-DAQ. The functionality of each thread varies based on the test category. 

The RTA employs both online and offline methods for data analysis. In online analysis every time the RTA receives data from the RPC-DAQ it promptly compares it with the generated configuration information. In offline analysis the RTA accumulates all RPC-DAQ data and stores it in a file. Once the test run is complete the RTA retrieves data from this file for analysis. Following the analysis the RTA generates a comprehensive test report regarding the Device Under Test (DUT) in this case the RPC-DAQ.

\section{Test Methodology}
\label{sec:Result}
RPC-DAQ uses various smaller functions that collectively work towards achieving primary goals. To ensure efficient performance of RPC-DAQs test methods are categorized into Primary and Secondary tests listed in Table~\ref{tab:1}. Primary tests involve assessing Event, Monitoring acquisition and cross talk in RPC-DAQ. Secondary tests has all peripheral level assessments such as Network performance, Pre-trigger generation, TDC logic, TPH sensor readout, Flash memory and RTC Test. If an RPC-DAQ successfully passes the primary tests there is no need to conduct any additional secondary functional tests. This is because the primary tests comprehensively cover the majority of RPC-DAQ functions. However, if an RPC-DAQ fails the primary tests subsequent secondary functional tests will be conducted. This paper specifically discusses the primary tests.

\begin{table}[htbp]
\centering
\caption{Classification of Test Methodologies.\label{tab:1}}
\smallskip
\begin{tabular}{l|l|l}
\hline
Category &Test Type &Description  \\
\hline
&Event Test  & RPC-DAQs capability of latching event data on a trigger\\
Primary &Monitoring Test & RPC-DAQs Detector Health Monitoring capablity \\
&Cross Talk Test & Study of Cross talk between channels\\
\hline
&Pretrigger Generation & Fold (Pretrigger) signals generation Functionality\\
Secondary&Timing Measurment & Testing the timing performance of RPC-DAQ and TDC\\
&Network Funtionality & Testing the network configuration and protocols\\
&RTC Synchronization & Local clock synchronization with Global clock \\
%$\alpha$ & $\beta$ & $\alpha$ and $\beta$\\
\hline
\end{tabular}
\end{table}

\subsection{Event Test}
\label{sec:cmd_be}
The primary function of RPC-DAQ is to capture position and timing information, facilitated by a dedicated feature known as Event Data Acquisition implemented in the RPC-DAQ FPGA. During event acquisition RPC-DAQ captures the position and the timing of the 128 incoming signals on a global trigger received from the trigger system. The captured hit latches and timing data along with the Real-Time Clock (RTC) timestamp are then framed into an event packet and transmitted to the data servers.

\begin{figure}[htbp]
\centering
\includegraphics[width=.7\textwidth]{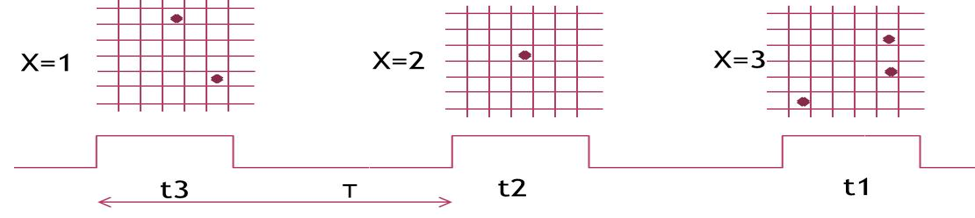}
\caption{Example of Event test pattern generated by RTA and testjig.\label{fig:8}}
\end{figure}

\begin{figure}[htbp]
\centering
\includegraphics[width=.8\textwidth]{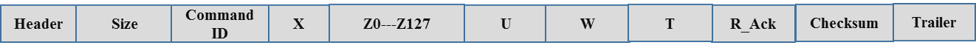}
\caption{Event test UDP Packet Format.\label{fig:9}}
\end{figure}

The goal of the Event Test is to collectively assess these functions. Upon receiving inputs from the user, the RTA generates a strip hit pattern as depicted in Figure~\ref{fig:8} and instructs the Test-Jig board to provide the RPC-DAQ with the generated input. Triggers are then sent to RPC-DAQ through the global services interface.  Figure~\ref{fig:9} illustrates the RTA command packet transmitted to the Test-Jig with event test pattern.

\begin{figure}[htbp]
\centering
\includegraphics[width=.8\textwidth]{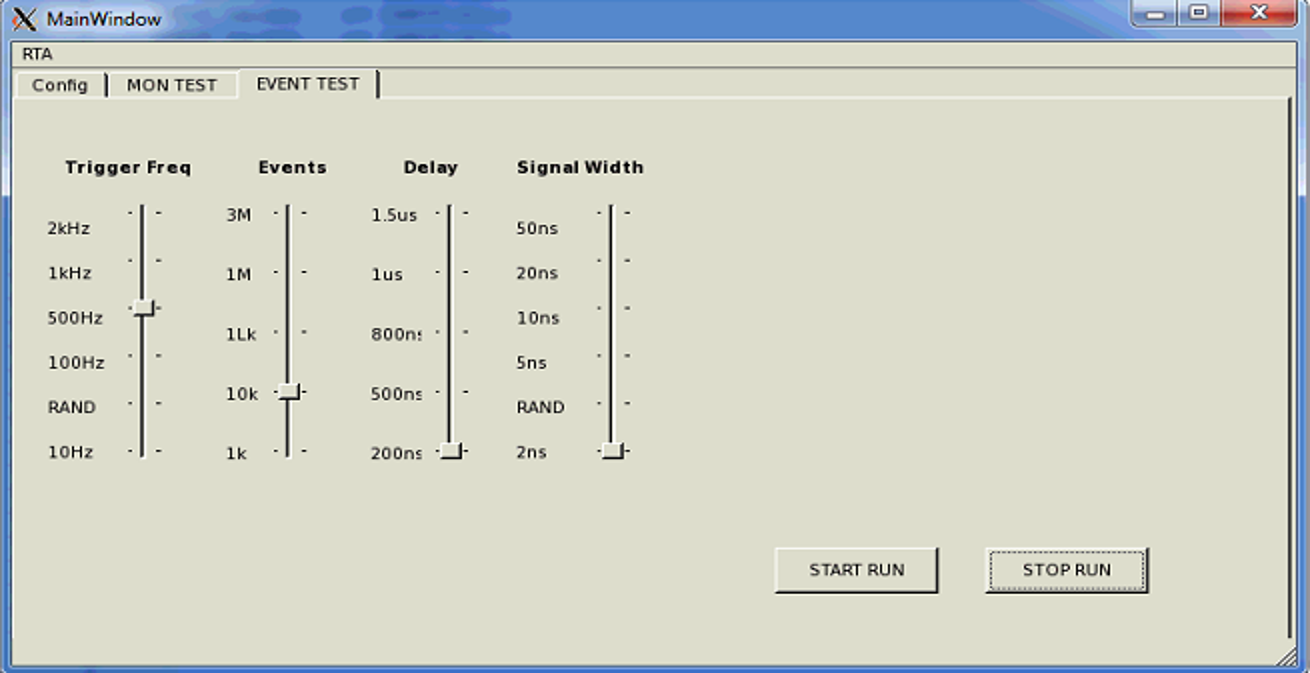}
\caption{Event Test method Graphical User Interface.\label{fig:10}}
\end{figure}

In the RTA Event test tab shown in Figure~\ref{fig:10} the user specifies (X) events to be generated at a trigger frequency rate (U), with a pre-defined strip hit pattern (Z0-Z127) retrieved from a file or randomly generated. The input signal width (W) is hard coded  and delay (T) between hit and trigger is also chosen. Utilizing this information, the RTA generates a test packet and dispatches it to the Test-Jig as a UDP command along with its header.
 
Upon receiving this packet the Test-Jig initiates the transmission of signals to RPC-DAQ. Whereas the RPC-DAQ captures these signals and generating an event packet for each event trigger and forwards them to the RTA back-end server. Upon receipt of the event packet from RPC-DAQ the RTA analyzes this data in comparison with the data it sent to the Test-Jig. This process iterates until the total number of events is reached. After the completion of the test the RTA generates a comprehensive report showcasing the performance and any identified errors.

\subsection{Monitoring Test}
\label{sec:cmd_fe}
\begin{figure}[htbp]
\centering
\includegraphics[width=.8\textwidth]{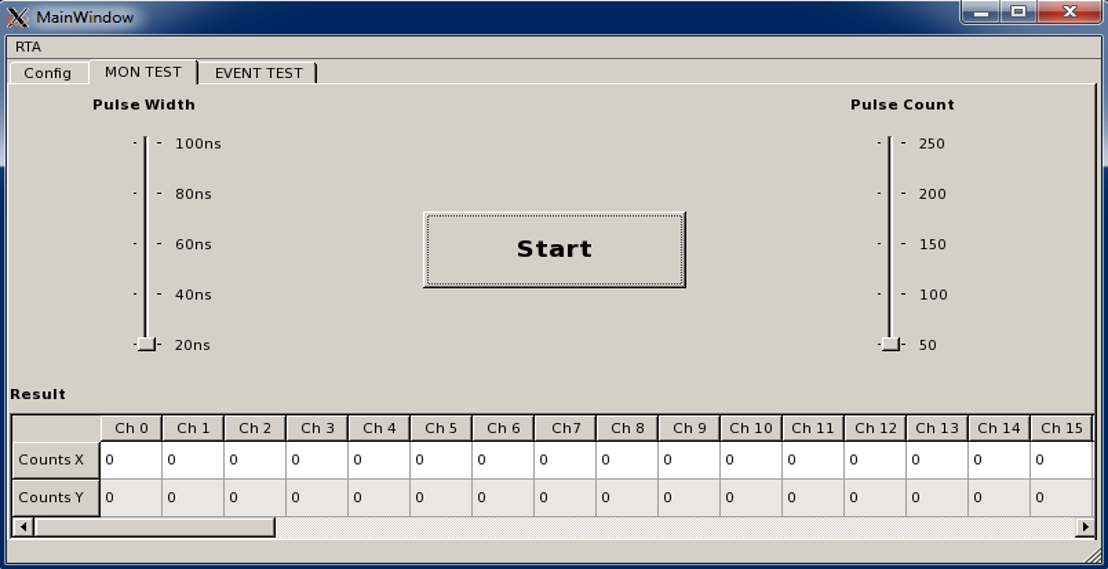}
\caption{MON Test method Graphical User Interface.\label{fig:12}}
\end{figure}

RPC-DAQ conducts periodic monitoring of detector health by counting the signals in all channels. This data is transmitted to an RPC Monitoring server for display and storage. Users access this information to adjust RPCs bias and ambient parameters. The Monitoring function in the RPC-DAQ FPGA counts the incoming pulses for each channel and sends this data along with the RTC timestamp to the Monitoring server.
\begin{figure}[htbp]
\centering
\includegraphics[width=.8\textwidth]{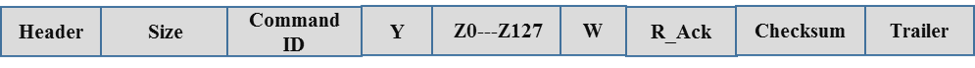}
\caption{MON Test UDP Packet Format.\label{fig:11}}
\end{figure}
In the Monitoring Test the RTA generates a monitoring command packet based on user input. This packet includes pulse count and pulse width information which is then sent to the Test-Jig. The Test-Jig in turn generates (W) pulses each with a width of (Y) for strips (Z). The command packet is illustrated in Figure~\ref{fig:11} , and the MON test GUI is depicted in Figure~\ref{fig:12} . The received data from the RPC-DAQ is displayed on the GUI and stored in a binary file for analysis. The analysis in this case compares the number of pulses generated in one channel with number of counts observed in a given monitoring interval.

\subsection{Cross Talk Test}
\label{sec:crc}
\begin{figure}[htbp]
\centering
\includegraphics[width=.8\textwidth]{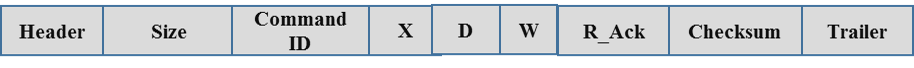}
\caption{Cross talk Test Packet.\label{fig:13}}
\end{figure}

The objective of this test is to verify that all 128 LVDS inputs of RPC-DAQ are free from cross talk. In this test methodology the Test-Jig generates input exclusively to a selected channel providing a finite amount (X) hits of input signals with a width (W) and a signal-to-trigger delay (D). The Test-Jig repeats this process in a loop for all other channels until the last channel is covered. 
The RTA receives data from RPC-DAQ and examines it for any cross talk between the selected channel and all the other channels. RTA analyzes each individual strip and informs users of any identified cross talk or the percentage of cross talk present. The RTA Cross talk command packet is presented in Figure~\ref{fig:13}.

\section{Test Results and Report Generation}
\label{sec:other}

After the completion of various tests the RTA prepares a test report that displays both the generated input and the observed measurements. This report is instrumental in studying the performance of the RPC-DAQ. The primary report consist of plots with measured timing and latch counts as functions of channels.  
% \begin{figure}[htbp]
% \centering
% \includegraphics[width=1\textwidth]{results.png}
% \caption{(Left) Event Latch test results, (Right) Leading and Trailing edge timing of channel 0.\label{fig:14}}
% \end{figure}

\begin{figure}[htbp]
\centering
\begin{subfigure}[c]{0.5\textwidth}
\includegraphics[width=\linewidth]{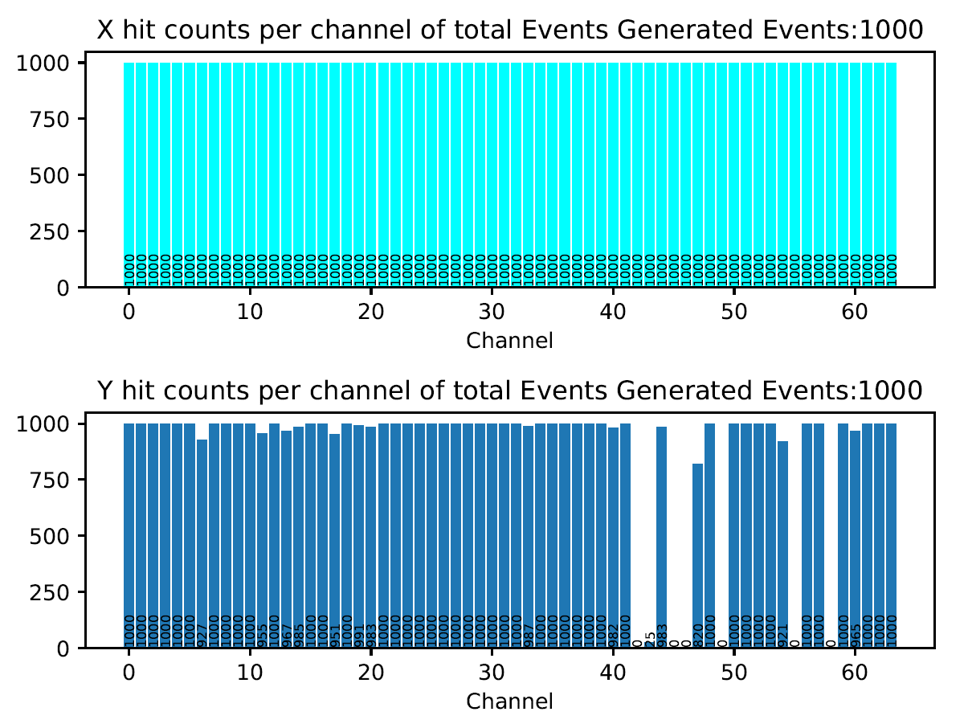} 
\caption{Event Latch test results.}
\label{fig:14a}
\end{subfigure}\hfill  
\begin{subfigure}[c]{0.5\textwidth}
\includegraphics[width=\linewidth]{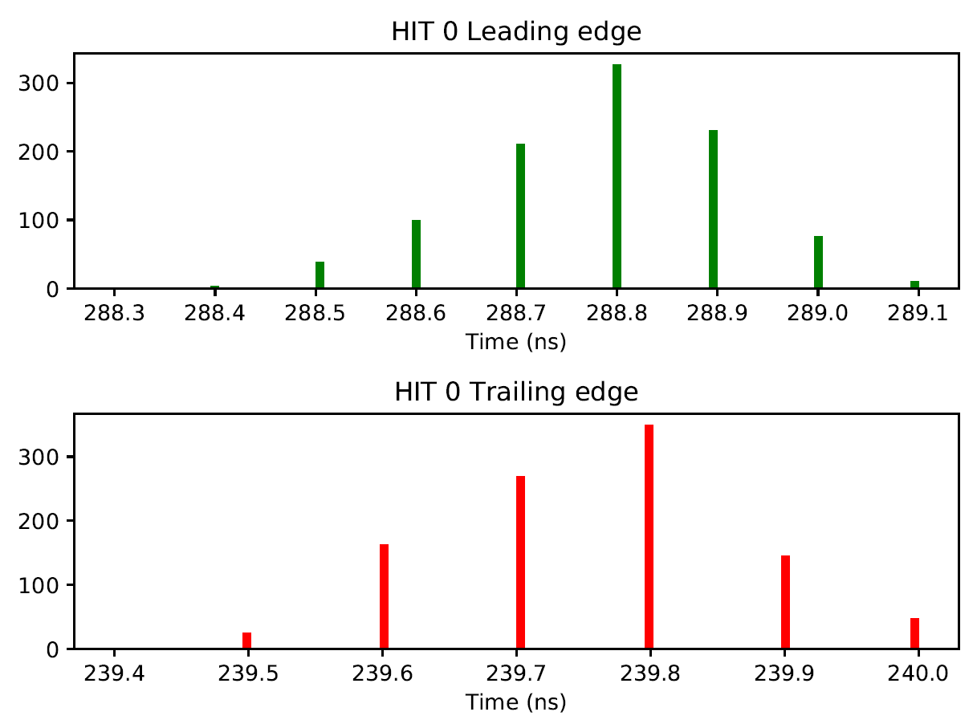}
\caption{Leading and Trailing edge timing of channel 0.}
\label{fig:14b}
\end{subfigure}
\caption{RPC-DAQ test results obtained using RTA.\label{fig:14}}
\end{figure}

As illustrated in the Figure~\ref{fig:14}(a) certain channels in the RPC-DAQ show lower counts compared to the global trigger count. This discrepancy can serve as an indicator to study the channel connectors for potential issues such as dry solder or broken pins. The timing plots on Figure~\ref{fig:14}(b) right side shows the signal quality of each channel. By examining these reports RTA users can evaluate the status of RPC-DAQ usage and potential areas of concern.

\section{Conclusion}

To verify all the functionalities of RPC-DAQ modules faster an automated system has been devised. This automated system integrates the Test-Jig module with application software known as RTA (RPC-DAQ Test Application). Through this system a range of test methodologies is implemented to comprehensively evaluate each function of the RPC-DAQ. The capability to test multiple RPC-DAQs simultaneously is still under development. Additionally, there is a planned comprehensive test, referred to as a Combined test which verifies all the functions of the RPC-DAQ under a unified test methodology.

% For internal references use label-refs: see section~\ref{sec:intro}.
% Bibliographic citations can be done with "cite": refs.~\cite{a,b,c}.

\acknowledgments
We sincerely thank all our present and former INO colleagues especially Pathaleswar, Dipanker Sil, Puneet Kanwar Kaur, Umesh L, Upendra Gokhale, Anand Lokapure, Suraj Kole, Sagar Sonavane, S.R. Joshi, K.C. Ravindran, Nagaraj Panyam, Rajkumar Bharathi for technical support during the design, Also we would like to thank members of TIFR, namely Piyush Verma, Darshana Gonji, Santosh Chavan and Vishal Asgolkar who supported testing and commissioning. Also we like to thank former INO directors N.K. Mondal and V. M Datar, for their continuous encouragement and guidance.  

% Pathaleswar, Dipanker Sil,  Puneet Kanwar Kaur, Umesh L, Upendra Gokhale,  Anand Lokapure, Pavan Kumar, Aditya Deodhar, Suraj Kole, Sagar Sonavane, Salam Thoi Thoi, Piyush Verma, Darshana Gonji, Santosh Chavan, Vishal Asgolkar

%Darshana Gonji, Santosh Chavan, Vishal Asgolkar, Piyush Verma,

\end{document}